\documentclass[aps,preprint,amsmath,amssymb,nofootinbib]{revtex4}

\usepackage{hyperref}
\usepackage{graphicx}
\usepackage{slashed}

\begin{document}

\title{Leptomeson contribution to the muon $\boldsymbol{g-2}$} 

\author{Dmitry Zhuridov}%
\email{dmitry.zhuridov@gmail.com}%

\affiliation{Institute of Physics, University of Silesia, Uniwersytecka 4, 40-007, Katowice, Poland}

\date{\today} 

\begin{abstract}
Many models on the market allow for particles carrying both lepton number and color, e.g., leptoquarks and leptogluons. 
Some of the models with this feature can also accommodate color-singlet \mbox{leptohadrons}. 
We have found that the long-standing discrepancy between the experimental result and 
the Standard Model prediction for the muon anomalous magnetic moment can be explained 
by the effect of leptomesons within the wide allowed range of their masses and couplings. 
These new particles are testable at the current run of the LHC.


\end{abstract}

\maketitle

\section{Introduction}
Nowadays physics at small scales is established to the characteristic distances for the nucleons substructure 
of about $10^{-15}$ m. There are many interesting theories which explore physics at even smaller distances, e.g., 
theories of extra dimensions and string theories. Yet another type of models constitute so-called composite 
models~\cite{Pati:1974yy,Terazawa:1976xx,Neeman:1979wp,Shupe:1979fv,Harari:1979gi,Squires:1980cm,Harari:1980ez,Fritzsch:1981zh,Barbieri:1981cy,King:1984vu}.    
The composite models---which contain colored preons---predict a possibility of leptoquarks 
and leptogluons. 
Leptoquarks are also present in many theories of grand unification~\cite{Pati:1974yy,Georgi:1974sy,Georgi:1974my}, 
R-parity-violating supersymmetric models~\cite{Barbieri:2005} 
and in the extended technicolor models~\cite{Lane:1991qh}. 
Many theories containing the particles that carry lepton number and color simultaneously 
predict color-singlet leptohadrons, in particular 
the bound states of leptoquark-antiquark pairs~\cite{Kuo:1997ax,Kiselev:1997si,BowserChao:1998wx}.  
However, their experimental signatures are poorly addressed in the literature. 
In this paper we consider the possible phenomenology of a particle that
interacts with a lepton and a meson and can be referred to as leptomeson (LM). 
Since LMs are colorless they can be lighter (and more accessible for collider probes) 
than the leptoquarks with their heavy color dressing. 

At low energies the new heavy particles may reveal themselves 
in the tiny effects on the lepton intrinsic properties (magnetic moment, charge radius, etc.). 
Currently one of the tools most sensitive to the new physics is the precise measurement of the muon 
anomalous magnetic moment (AMM) $a_\mu = (g_\mu-2)/2$, which reveals a \mbox{$3$--$4~\sigma$} discrepancy 
with its theoretical prediction within the Standard Model (SM)~\cite{Jegerlehner:2009ry,Hagiwara:2011af,PDG2014}. 
In the theories of compositeness the new contributions to the lepton AMM can arise from 
the relativistic bound state description of leptons~\cite{Brodsky:1980zm,Dai:2001vv}. 
Among the possible new heavy composites, purely leptogluons may contribute 
to the lepton AMM starting only from the two-loop level. 
However, one-loop contributions can be generated by leptogluons plus color-octet $Z$ bosons~\cite{Das:2001it}, 
the colorless excited leptons~\cite{Renard:1982ij,Mery:1989dx}, and 
leptoquarks~\cite{Djouadi:1989md,Chakraverty:2001yg,Cheung:2001ip,Biggio:2014ela}. 

Leptomesons (leptodiquarks) may effectively provide a new one-loop (two-loop) contribution to the muon AMM. 
A not too large size for this contribution 
can be achieved due to large LM masses of $\mathcal{O}(100)$~GeV and/or small LM 
couplings, e.g., of $\mathcal{O}(10^{-2})$. 

Many kinds of one-loop contributions to $a_\mu$ were classified~\cite{Freitas:2014pua,Queiroz:2014zfa} 
by the types of particles that propagate in the loop, 
which typically include light SM fermions (leptons and light quarks) and heavy vectors, scalars and fermions 
($W$, $Z$, $H$, $t$, and new heavy particles). 
In addition, there is a possibility of very light [with the masses of $\mathcal{O}(\text{1})$~GeV] 
superweakly coupled new particles~\cite{Gninenko:2001hx,Fayet:2007ua,Pospelov:2008zw,Davoudiasl:2012ig,Chen:2015vqy} 
propagating in the loop, which is hard to test in collider experiments. 
However, the discussed leptomeson contribution to $a_\mu$ may involve light 
vector and/or scalar particles (vector and scalar mesons), and can be tested at colliders. 
%

In this paper we investigate the effects of LMs on the muon $g-2$. 
To our knowledge, these effects have not been considered yet in the literature. 
In the next section we give analytical expressions for the characteristic LM contributions to $a_\mu$. In section~\ref{sec:results} we 
present and discuss the numerical results, and we conclude in section~\ref{sec:conclusion}.

\section{Leptomesons in muon $\boldsymbol{g-2}$}\label{section:LM}

The present discrepancy between the experimental data and the SM calculation for the muon 
AMM is~\cite{Hagiwara:2011af}
\begin{eqnarray}\label{Eq:a_mu}
     \Delta a_\mu = 26.1 (8.0) \times 10^{-10},
\end{eqnarray}
which is 3.3~$\sigma$. Recent progress in calculating the hadronic contribution increases 
this discrepancy up to 4.0 $\sigma$~\cite{Jegerlehner:2015stw}. 
A further essential increase is possible after more precise measurements of $a_\mu$ in 
the near-future experiments at Fermilab~\cite{Grange:2015eea} and J-PARC~\cite{Saito:2012zz}.
We consider the possibilities of explaining this anomaly using LM contributions.

For each lepton flavor one can expect several neutral LMs 
$\ell^0_{Si}$ ($\ell^0_{Vi}$), which interact with the mesons of scalar (vector) type $S$ ($V$) 
and decay into the lepton-meson pairs: $\ell^+S^-$, $\ell^-S^+$, $\nu_\ell S^0$ 
($\ell^+V^-$, $\ell^-V^+$, $\nu_\ell V^0$). 
Concerning the charged LMs, we presume the singly charged $\ell_{S(V)}^\pm$ and 
the doubly charged $\ell_{S(V)}^{\pm\pm}$ states, 
which can decay into \{$\ell^\pm S^0$, $\nu_\ell S^\pm$\} (\{$\ell^\pm V^0$, $\nu_\ell V^\pm$\}) 
and $\ell^\pm S^\pm$ ($\ell^\pm V^\pm$), respectively.\footnote{To be definite we classify 
LMs by the type of their interaction with mesons (either scalar or vector). 
However, in general, one LM may have both scalar and vector type interactions.}
Due to the large number of mesons and possible variety of LMs we 
restrict our consideration by typical scalar and vector meson-LM-lepton interactions, which may 
give significant corrections to 
the lepton AMM $a_\ell$. 

The lowest-dimension Lagrangian for the effective interactions of neutral, singly charged and 
doubly charged LMs with the charged leptons and the mesons can be written as 
\begin{eqnarray}\label{Eq:Lagrangian}
    \mathcal{L}_\text{LM} &=& -\sum_{\alpha,\beta=L,R \,(\alpha\neq\beta)} 
    \bar\ell_\beta (g_{\alpha S}^0\ell_{\alpha S}^0 \,S^- + g_{\alpha S}^-\ell_{\alpha S}^- \,S^0 
    							+ g_{\alpha S}^{-\,-}\ell_{\alpha S}^{-\,-} \,S^+)    \nonumber\\
		  &-&\sum_{\alpha=L,R} \bar\ell_\alpha \gamma^\mu (g_{\alpha V}^0\ell_{\alpha V}^0 \,V_\mu^- + g_{\alpha V}^-\ell_{\alpha V}^- \,V_\mu^0 
		  					+ g_{\alpha V}^{-\,-}\ell_{\alpha V}^{-\,-} \,V_\mu^+)    + \text{H.c.},
\end{eqnarray} 
where 
$\ell=e,\mu,\tau$ is the charged lepton, 
and $g_{\alpha S}$ and $g_{\alpha V}$ are the new dimensionless couplings with suppressed flavor indices. 
For simplicity we require real couplings (to avoid constraints from  the  electric  dipole  moment 
of the electron) and flavor conserving interactions in Eq.~\eqref{Eq:Lagrangian}. 
%
\begin{center}
 \begin{figure}[tb]
 \centering
 	\includegraphics[width=0.3\textwidth]{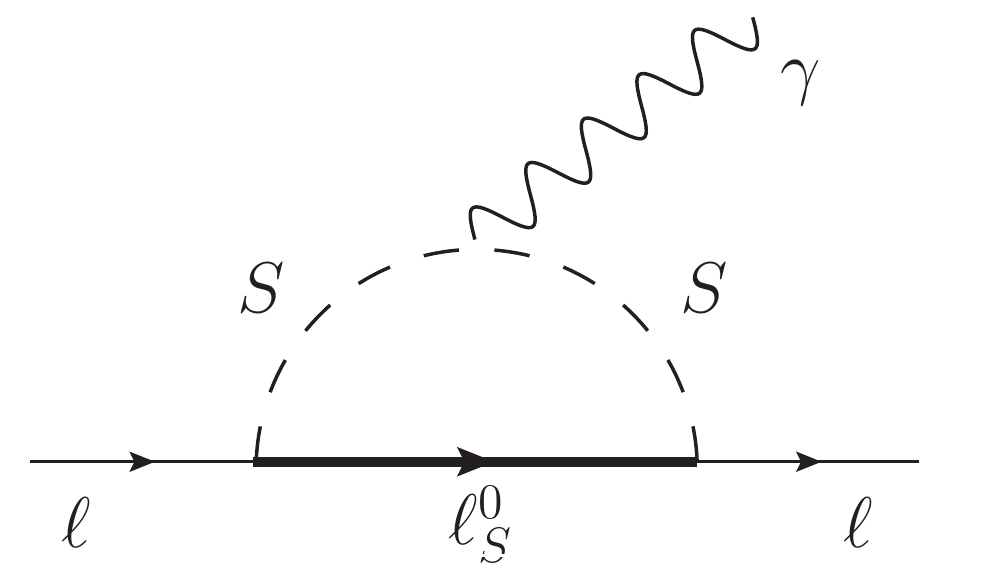} \ 
 	\includegraphics[width=0.3\textwidth]{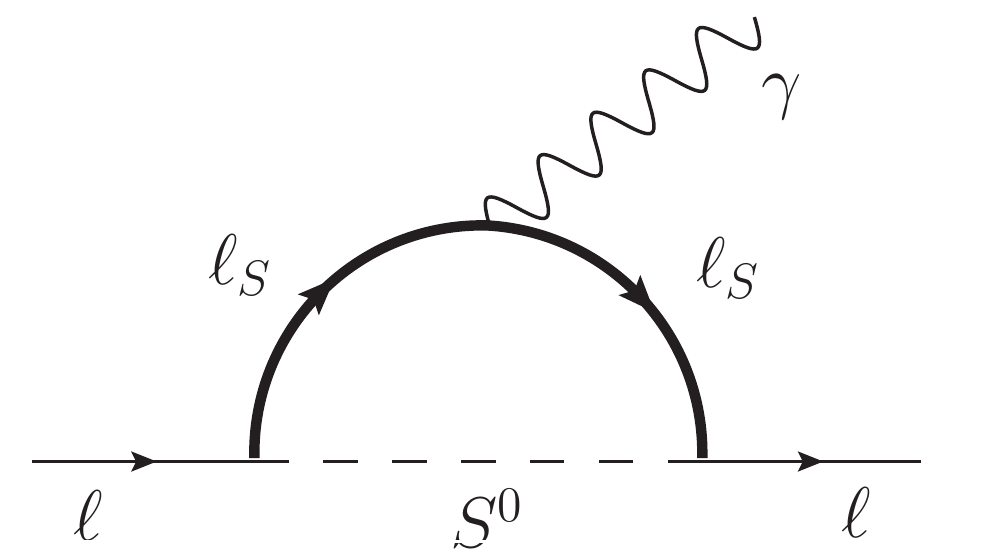}
	   \caption{Feynman diagrams for the one-loop contribution to $g_\ell-2$ for 
	   the leptomesons: neutral $\ell_S^0$ (left) and singly charged $\ell_S$ (right). 
	 }
   \label{Fig:diagrams1}
 \end{figure}
\end{center}
\begin{center}
 \begin{figure}[tb]
 \centering
 	\includegraphics[width=0.3\textwidth]{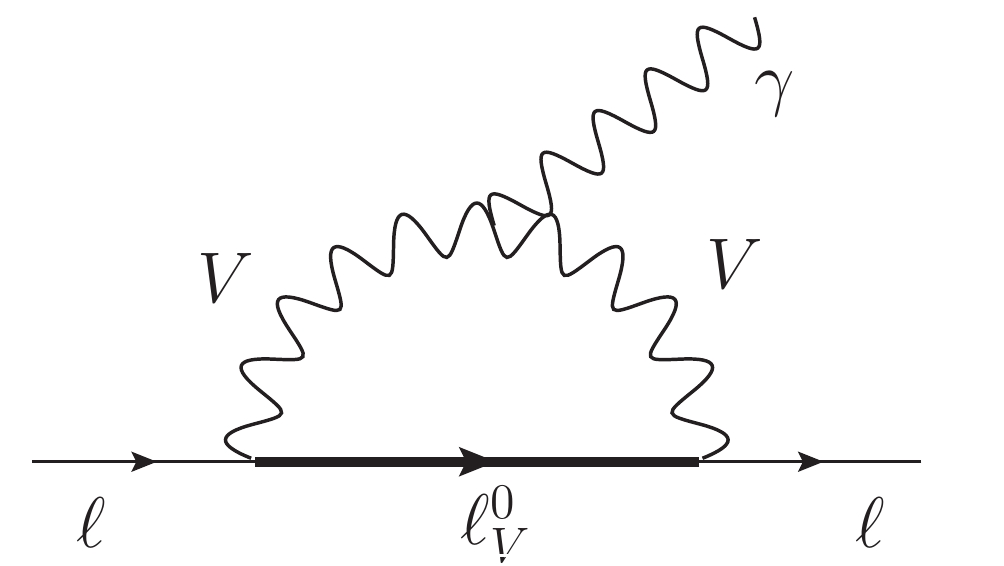} \ 
 	\includegraphics[width=0.3\textwidth]{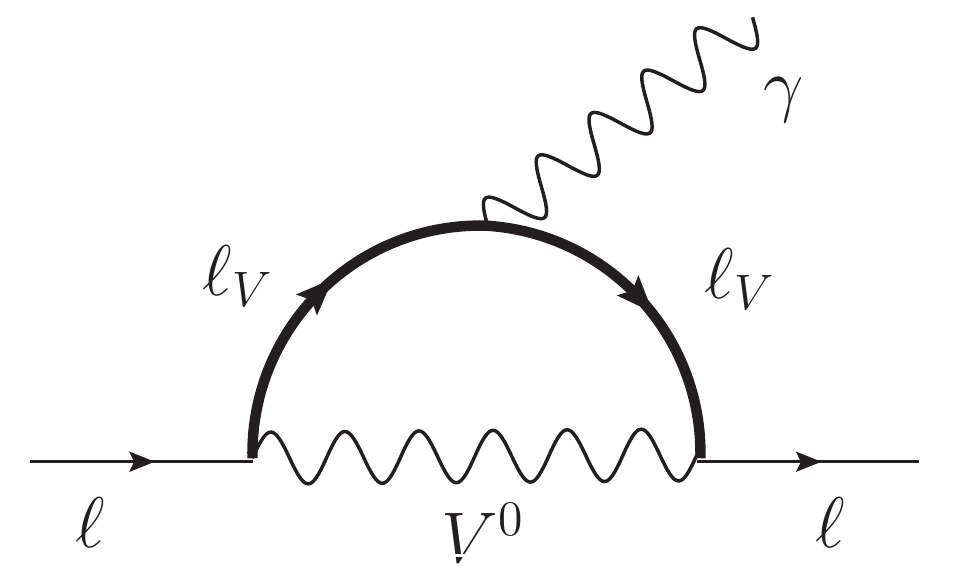}
	   \caption{Feynman diagrams for the one-loop contribution to $g_\ell-2$ for 
	   the leptomesons: neutral $\ell_V^0$ (left) and singly charged $\ell_V$ (right). 
	 }
   \label{Fig:diagrams2}
 \end{figure}
\end{center}

The leading contributions to the lepton AMM from the one-loop processes with neutral and singly charged 
scalar and vector LMs are represented in Figs.~\ref{Fig:diagrams1} and \ref{Fig:diagrams2}.\footnote{Contributions from 
these diagrams compete with the two-loop contributions with propagating leptodiquark and two quarks.}
Using the generic analytic formulas for the one-loop contributions to the lepton 
AMM~\cite{Vendramin:1988rd,Freitas:2014pua}  
in the limit of large LM masses with respect to the lepton mass $m_\ell$ and the meson masses,  
these contributions can be written as
\begin{eqnarray}
\label{Eq:AB_S_approx}
    \Delta a_{\ell_S^0} &=&  -\, \frac{(g_{\alpha S}^{0})^2}{16\pi^2}\, \frac{m_\ell^2}{M_0^2} 
	 \, f_\text{SSF}\left( \frac{m_{S^\pm}^2}{M_0^2} \right),  \\
\label{Eq:AB_S0_approx}
    \Delta a_{\ell_S^-} &=& \quad \frac{(g_{\alpha S}^{-})^2}{16\pi^2}\, \frac{m_\ell^2}{M^2} 
	 \, f_\text{FFS}\left( \frac{m_{S^0}^2}{M^2} \right), \\
\label{Eq:AB_V0_approx}
    \Delta a_{\ell_V^0} &=& \quad  \frac{(g^{0}_{\alpha V})^2}{16\pi^2}\, \frac{m_\ell^2}{m_{V^\pm}^2}  \, 
     f_\text{VVF}\left( \frac{m_{V^\pm}^2}{M_0^2} \right), \\
\label{Eq:AB_V_approx}
    \Delta a_{\ell_V^-} &=&  -\, \frac{(g^{-}_{\alpha V})^2}{16\pi^2}\, \frac{m_\ell^2}{m_{V^0}^2}  \, 
     f_\text{FFV}\left( \frac{m_{V^0}^2}{M^2} \right),
\end{eqnarray}
where $m_S$ ($m_V$) is the scalar (vector) meson mass, 
$M$ ($M_0$) is the mass of the charged (neutral) LM, and the loop functions are
\begin{eqnarray}
    f_\text{SSF}(x)  &=&  \frac{1}{6(1-x)^4} [2 + 3x - 6x^2 + x^3 + 6x\,\text{ln}\,x]  \nonumber\\
    &\approx&   \frac{1}{3} + \frac{11}{6}x + x\,\text{ln}\,x, 
    \label{Eq:tildA_S} \\
    f_\text{FFS}(x)  &=&  \frac{1}{6(1-x)^4} [1 - 6x + 3x^2 + 2x^3 - 6x^2\,\text{ln}\,x]  \nonumber\\
    &\approx&   \frac{1}{6} - \frac{x}{3}, 
    \label{Eq:tildA_S0}  \\
    f_\text{VVF}(x)  &=&  \frac{1}{6(1-x)^4} [4 - 49x + 78x^2 - 43x^3 + 10x^4 - 18x\,\text{ln}\,x]  \nonumber\\
    &\approx&   \frac{2}{3} - \frac{11}{2}x - 3x\,\text{ln}\,x, 
    \\
    f_\text{FFV}(x)  &=&  \frac{1}{6(1-x)^4} [5 - 14x + 39x^2 - 38x^3 + 8x^4 + 18x^2\,\text{ln}\,x]  \nonumber\\
    &\approx&   \frac{5}{6} + x, 
    \label{eq:FFV}
\end{eqnarray}
where in the approximate expressions we neglected the terms of $\mathcal{O}(x^2)$. 
Clearly, these functions are positive for small $x$, and 
the contributions in Eqs.~\eqref{Eq:AB_S_approx} and \eqref{Eq:AB_V0_approx} are negative, 
while in Eqs.~\eqref{Eq:AB_S0_approx} and \eqref{Eq:AB_V_approx} they are positive. 
The relations to the loop functions given in Eq.~(A.1) 
of Ref.~\cite{Freitas:2014pua} are as follows
\begin{eqnarray}
    f_\text{SSF}(x)  &\equiv&  - \frac{1}{x} F_\text{SSF}\left(\frac{1}{x}\right) = F_\text{FFS}(x),  \\
    f_\text{FFS}(x)  &\equiv&    \frac{1}{x} F_\text{FFS}\left(\frac{1}{x}\right) = - F_\text{SSF}(x),  \\
    f_\text{VVF}(x)  &\equiv&    F_\text{VVF}\left(\frac{1}{x}\right),  \\
    f_\text{FFV}(x)  &\equiv&  - F_\text{FFV}\left(\frac{1}{x}\right).
\end{eqnarray}

The contributions of the scalar LMs in Eqs.~\eqref{Eq:AB_S_approx} and \eqref{Eq:AB_S0_approx} are suppressed 
by the second power of the meson-to-LM mass ratio. The case of vector LMs is special since their contributions 
in Eqs.~\eqref{Eq:AB_V0_approx} and \eqref{Eq:AB_V_approx} do not have this suppression. 
Hence smaller values of the couplings are required 
for the vector LM interactions to not to exceed the discrepancy in Eq.~\eqref{Eq:a_mu}. 

Notice that in the case of one-loop contributions to $a_\ell$ of a scalar $S$ ($g_{LS}=g_{RS}$) and a pseudoscalar 
$P$ ($g_{LS}=-g_{RS}$) the lepton mass $m_\ell$ receives large loop corrections unless the chirally 
symmetric limit $m_S=m_P$ is satisfied~\cite{Marciano,Czarnecki:2000id}.

\begin{center}
 \begin{figure}[tb]
 \centering
 	\includegraphics[width=0.3\textwidth]{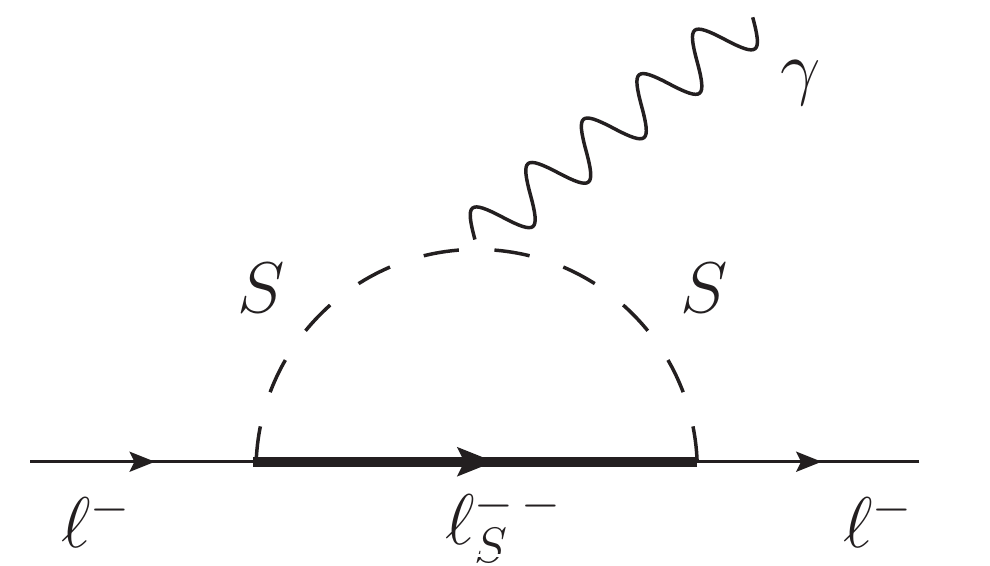} \ 
 	\includegraphics[width=0.3\textwidth]{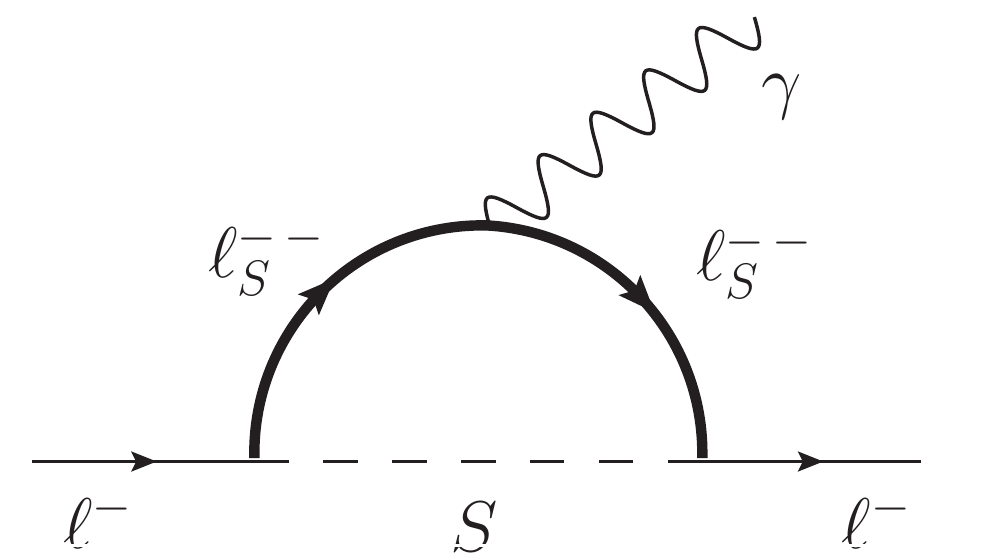} \\
 	\includegraphics[width=0.3\textwidth]{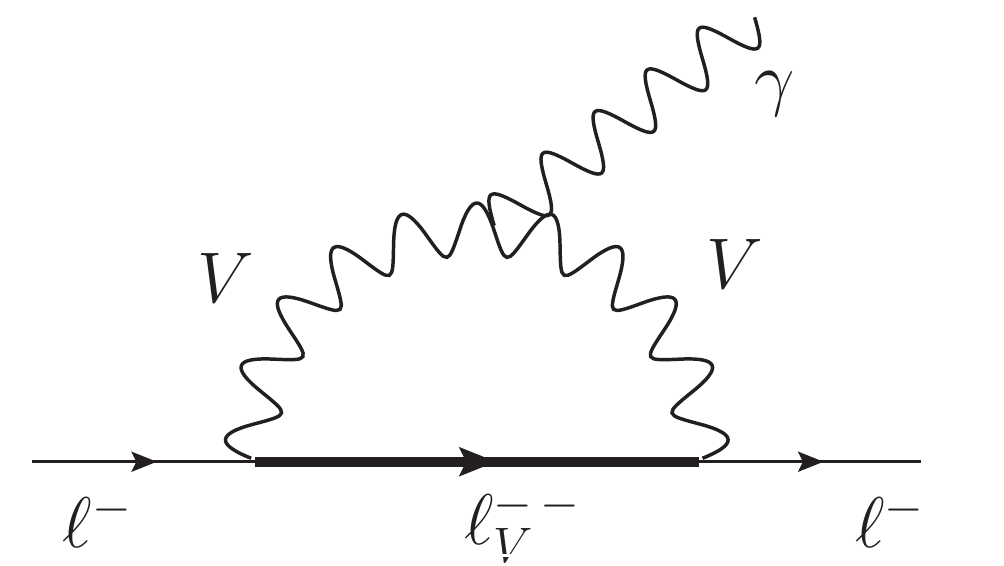} \
 	\includegraphics[width=0.3\textwidth]{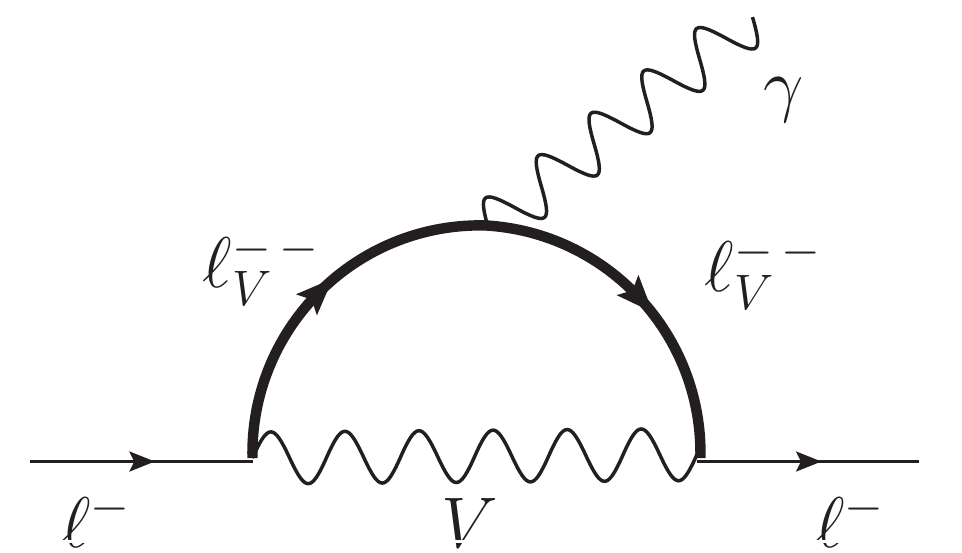}
	   \caption{Feynman diagrams for the one-loop contribution to the AMM of $\ell^-$ for 
	   the doubly charged leptomesons: $\ell_S^{-\,-}$ (upper) and $\ell_V^{-\,-}$ (lower). 
	 }
   \label{Fig:diagrams3}
 \end{figure}
\end{center}
The leading contributions to the lepton AMM from the one-loop processes with 
doubly charged scalar and vector LMs are shown in Fig.~\ref{Fig:diagrams3}, and can be written as
\begin{eqnarray}
    \Delta a_{\ell_S^{-\,-}} &=&  -\, \frac{(g_{\alpha S}^{-\,-})^2}{16\pi^2}\, \frac{m_\ell^2}{M_{-\,-}^2} 
	 \, f_\text{SF}\left( \frac{m_{S^\pm}^2}{M_{-\,-}^2} \right),  \\
%
    \Delta a_{\ell_V^{-\,-}} &=&  -\, \frac{(g^{-\,-}_{\alpha V})^2}{16\pi^2}\, \frac{m_\ell^2}{m_{V^\pm}^2}  \, 
     f_\text{VF}\left( \frac{m_{V^\pm}^2}{M_{-\,-}^2} \right),
\end{eqnarray}
where $M_{--}$ is the mass of the doubly charged LM, and the loop functions are
\begin{eqnarray}
    f_\text{SF}(x)  &=&  f_\text{SSF}(x) - 2f_\text{FFS}(x)  \label{Eq:fSF}\\
    &=&  \frac{x}{2(1-x)^4} [5 - 4x - x^2 + (2+4x)\,\text{ln}\,x]  \nonumber\\
    &=&  \frac{5}{2}x + x\,\text{ln}\,x  +  \mathcal{O}(x^2),\nonumber\\ 
    f_\text{VF}(x)  &=&  -f_\text{VVF}(x) + 2f_\text{FFV}(x) \label{Eq:fVF}\\
    &=&  \frac{1+2x}{2(1-x)^4} [2 + 3x - 6x^2 + x^3 + 6x\,\text{ln}\,x]  \nonumber\\
    &=& 1 + \frac{15}{2}x + 3x\,\text{ln}\,x  +  \mathcal{O}(x^2),\nonumber
\end{eqnarray}
where the factors of 2 in Eqs.~\eqref{Eq:fSF} and \eqref{Eq:fVF} come from the fact that the 
electromagnetic interaction of the doubly charged LM is twice as strong as that of a singly charged particle. 
Both scalar and vector doubly charged LM contributions are negative.

We demonstrate in section~\ref{sec:results} that a single {\it mumeson} (muonic leptomeson)---which is either a 
charged scalar $\mu_S^-$ or 
neutral vector $\mu_V^0$, and dominantly interacts with a specific meson---can provide the observed value of $a_\mu$ 
for the mass of a few hundred GeV with the couplings of 
$g_{\alpha S}^-=\mathcal{O}(1)$ and $g_{\alpha V}^0=\mathcal{O}(10^{-2})$, respectively. 
The case with either several mumesons or one mumeson that has a significant interaction with several mesons, 
which essentially contribute to $a_\mu$, is more involved and potentially has richer phenomenology. 
We discuss this case in several examples.

\section{Results and Discussion}\label{sec:results}

The new particles in various theories can be generically constrained using 
the parameters $S$ and $T$~\cite{Peskin:1991sw}. 
However, large contributions to the $T$ parameter are excluded in the case of approximate  mass  degeneracy  
of the components of the new weak multiplets, while a significant change of the $S$ parameter  
is avoided in case of vector-like couplings of new fermions to the gauge bosons 
(this simultaneously ensures the cancellation of axial-vector gauge anomalies)~\cite{Freitas:2014pua}.
The existence of the three generations of the SM leptons supposes analogous generations 
of the leptohadrons. This allows one to accommodate the assumption of minimal flavor violation~\cite{D'Ambrosio:2002ex}, 
which may help to avoid the constraints from the nonobservation of the flavor-violating processes 
(such as $\mu\to e\gamma$) and to protect the muon mass from large corrections 
induced by vector fermions~\cite{Freitas:2014pua}.
%
The nonobservation of new fermions in $e^+e^-$ collisions at the LEP collider at center-of-mass energies 
$\sqrt{s}\approx 200$~GeV yields a generic lower bound on the LM mass of $M\gtrsim100$~GeV. 
Moreover $Z$-pole precision measurements at LEP strongly constrain the vector LM mixings. 

\begin{center}
 \begin{figure}[tb]
 \centering
 	\includegraphics[width=0.45\textwidth]{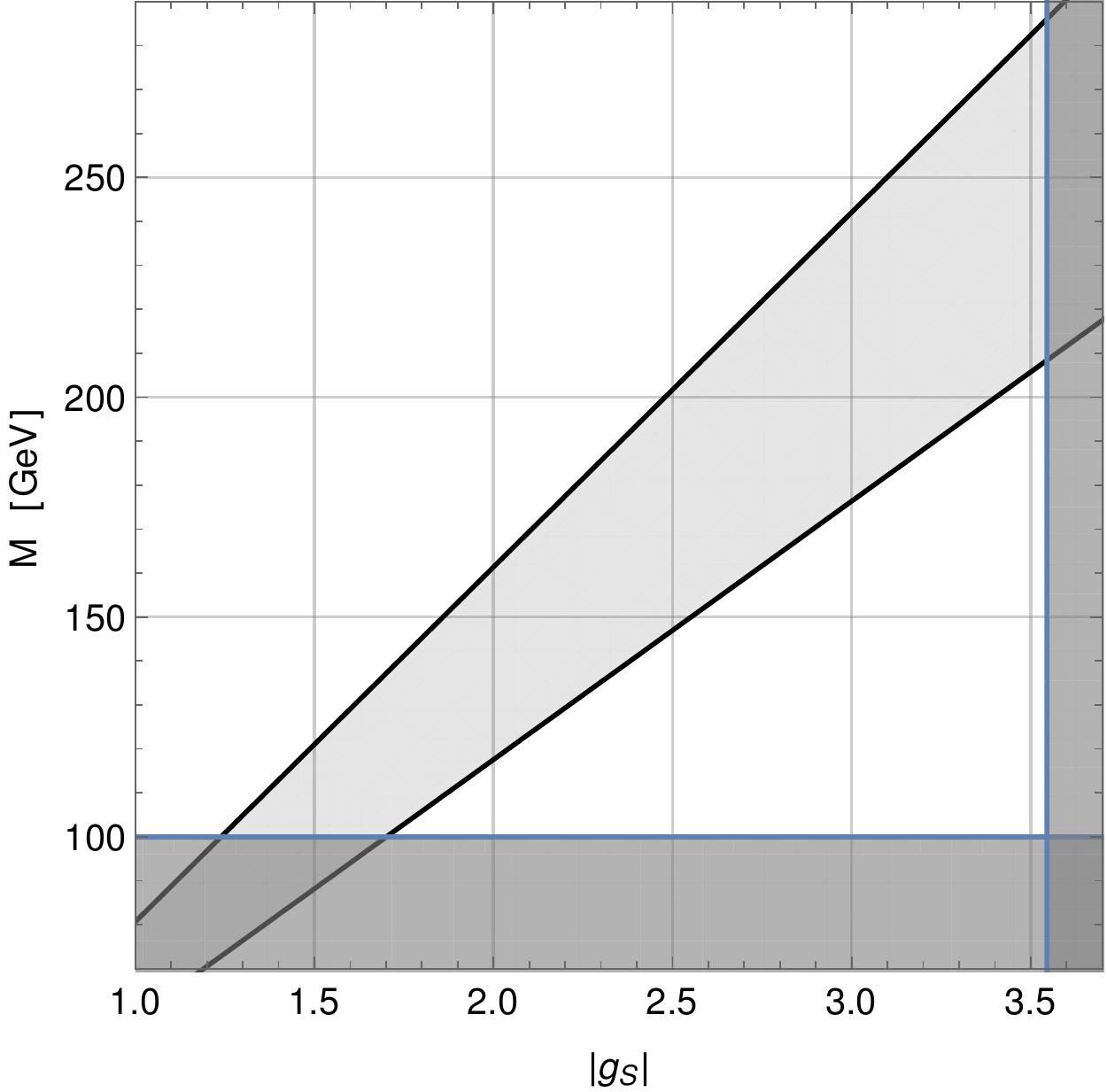} \quad 
 	\includegraphics[width=0.45\textwidth]{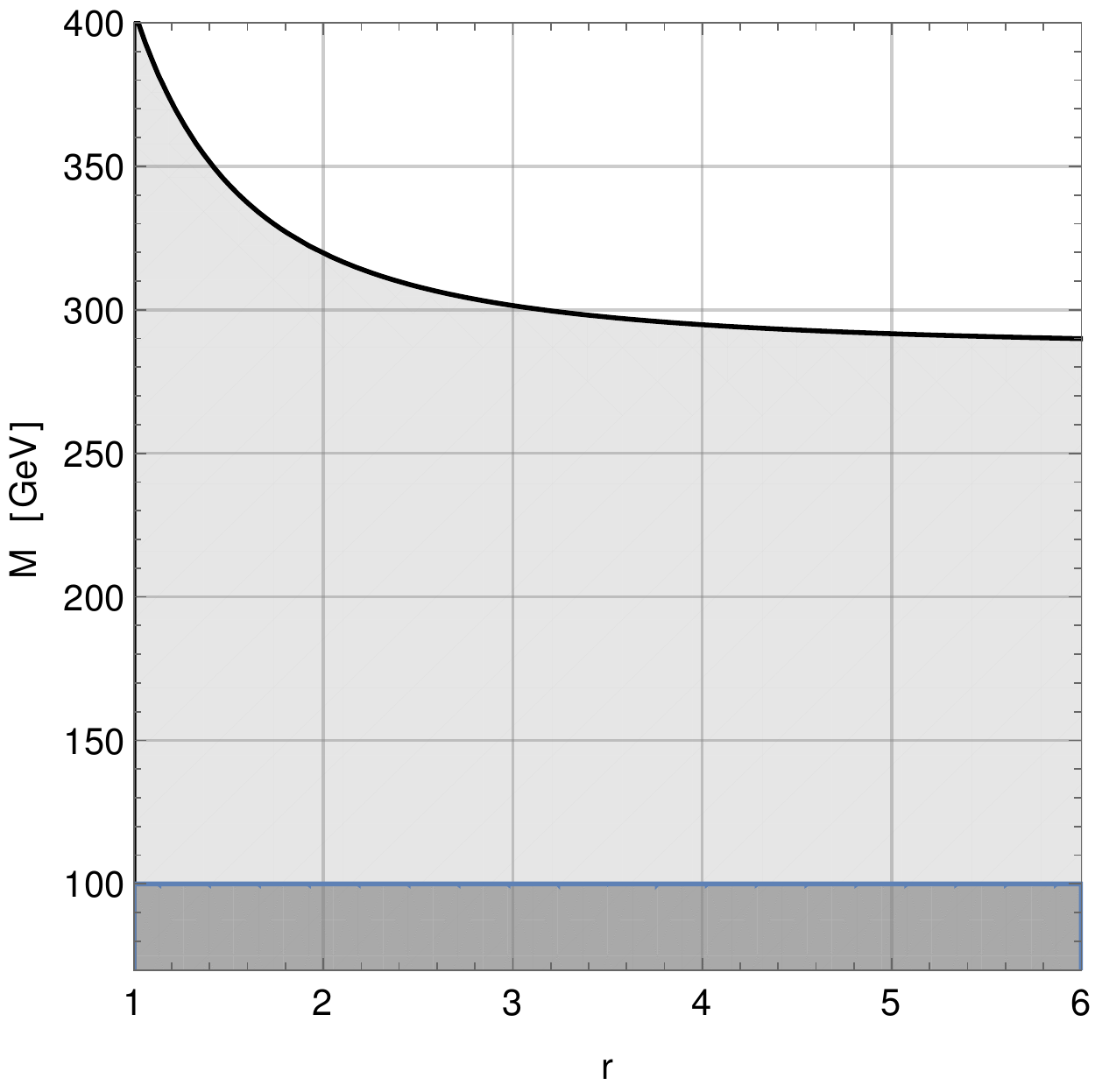}
	   \caption{\label{Fig:g_M} Allowed regions for the charged mumeson(s) interacting with the scalar meson(s).\\
	  {\it Left}: Mumeson mass $M$ vs coupling $|g_S|$ for a single meson-mumeson-lepton interaction. 
	   The light grey region is allowed by the current 
	   $(g-2)_\mu$ data within the 1\,$\sigma$ range.  
	   The dark grey regions represent the generic LEP mass bound of $M>100$~GeV 
	   and the perturbativity bound of $|g_S|<\sqrt{4\pi}$. 
	  {\it Right}: Mass $M$ vs ratio $r$ in case of two mumeson contributions to $a_\mu$. 
	   The light grey region is allowed by the current $a_\mu$ data (within 1\,$\sigma$) 
	   and the perturbativity bound of $|g_{Si}|<\sqrt{4\pi}$.  
	   The dark grey region is disfavored by the LEP data.
	 }
   \label{Figdistr}
 \end{figure}
\end{center}

\begin{center}
 \begin{figure}[tb]
 \centering 
 	\includegraphics[width=0.3\textwidth]{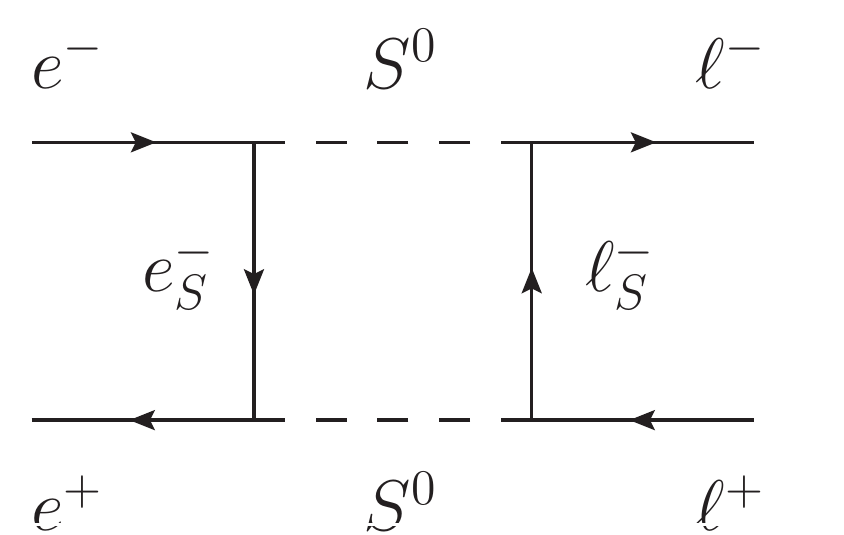} \quad 
 	\includegraphics[width=0.3\textwidth]{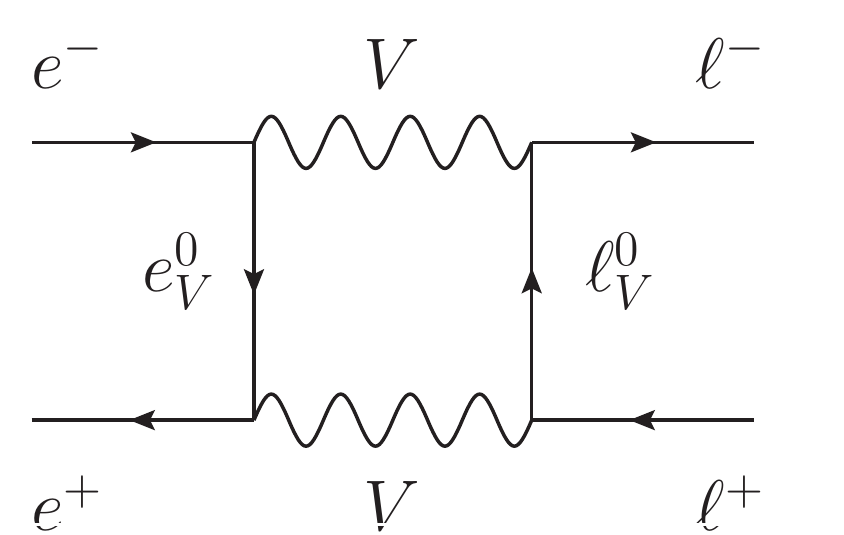}
	   \caption{\label{Fig:diagram1}One-loop leptomeson $\ell_S$ (left) and $\ell^0_V$ (right) 
	   contributions to the effective four-lepton interactions $ee\ell\ell$. 
	 }
   \label{Fig:box_V}
 \end{figure}
\end{center}
\subsection{Scalar leptomeson contribution}

For scalar mumesons the value of $a_\mu$ is insensitive to the meson masses for $m_{S^0}\ll M$. 
The proper value of $a_\mu$ can be provided for a single charged mumeson $\mu_S^-$  with a mass $M$ 
below 280~GeV as shown in Fig.~\ref{Fig:g_M} (left). In this figure we assume a significant effect 
on $a_\mu$ of only one coupling $g_S\equiv g_{\alpha S}^-$, where $\alpha$ is either $R$ or $L$. 
It is clear that small values of the coupling $|g_S|$ below 1.2 are ruled out.
Constraints from four-lepton contact interactions are absent if the meson $S^0$ 
is self-conjugate since the box diagram shown in Fig.~\ref{Fig:box_V} (left) is canceled by
a second diagram with crossed fermion lines in the final state.

In the case of two significant scalar mumeson contributions to $a_\mu$, 
the lightest mumeson mass $M$ can be as large as 400~GeV, which is shown in Fig.~\ref{Fig:g_M} (right) for 
the charged mumesons. In this figure the ratio $r$ is defined as
\begin{eqnarray}
  r = \frac{M_2}{M_1} \frac{|g_{S1}|}{|g_{S2}|} < \frac{M_2}{M_1} \frac{\sqrt{4\pi}}{|g_{S2}|},
\end{eqnarray}
where $M_i$ ($i=1,2$; $M\equiv M_1\leq M_2$) and $g_{Si}\equiv g^-_{\alpha Si}$ ($|g_{S1}|\geq |g_{S2}|$) 
are the mumeson masses and couplings, respectively. 
In particular, in the case of two scalar mumesons $\mu_{S1}$ and $\mu_{S2}$
that interact with a meson $S^0$ with the same coupling this ratio is reduced to $r=M_2/M_1$, 
while in the case of one scalar mumeson, which significantly interacts with two mesons 
$S_1^0$ and $S_2^0$ with different couplings $g_{S1}\neq g_{S2}$, 
the ratio is reduced to $r=|g_{S1}/g_{S2}|<4\pi/|g_{S2}|$.

The lower bound on $\mu_S^\pm$ mass can be significantly increased through the searches 
for their pair production at the LHC. However in this case the final state is composed of a dimuon and neutral 
meson-antimeson pair (the decay of which may give photons, leptons, $\pi^\pm$, etc.) 
instead of the final state of dilepton plus either gluonic or quark jets, which was considered in 
the searches for leptoquarks~\cite{CMS:2014qpa,Aad:2015caa} 
and leptogluons~\cite{Goncalves-Netto:2013nla,Jelinski:2015epa,Acar:2015wxp}. 
Another alternatives include searches for single $\mu_S^\pm$ productions, and 
$\mu^+\mu^-$ production in meson-meson fusion via $t$-channel exchange of $\mu_S^\pm$. 

\begin{center}
 \begin{figure}[tb]
 \centering
 	\includegraphics[width=0.45\textwidth]{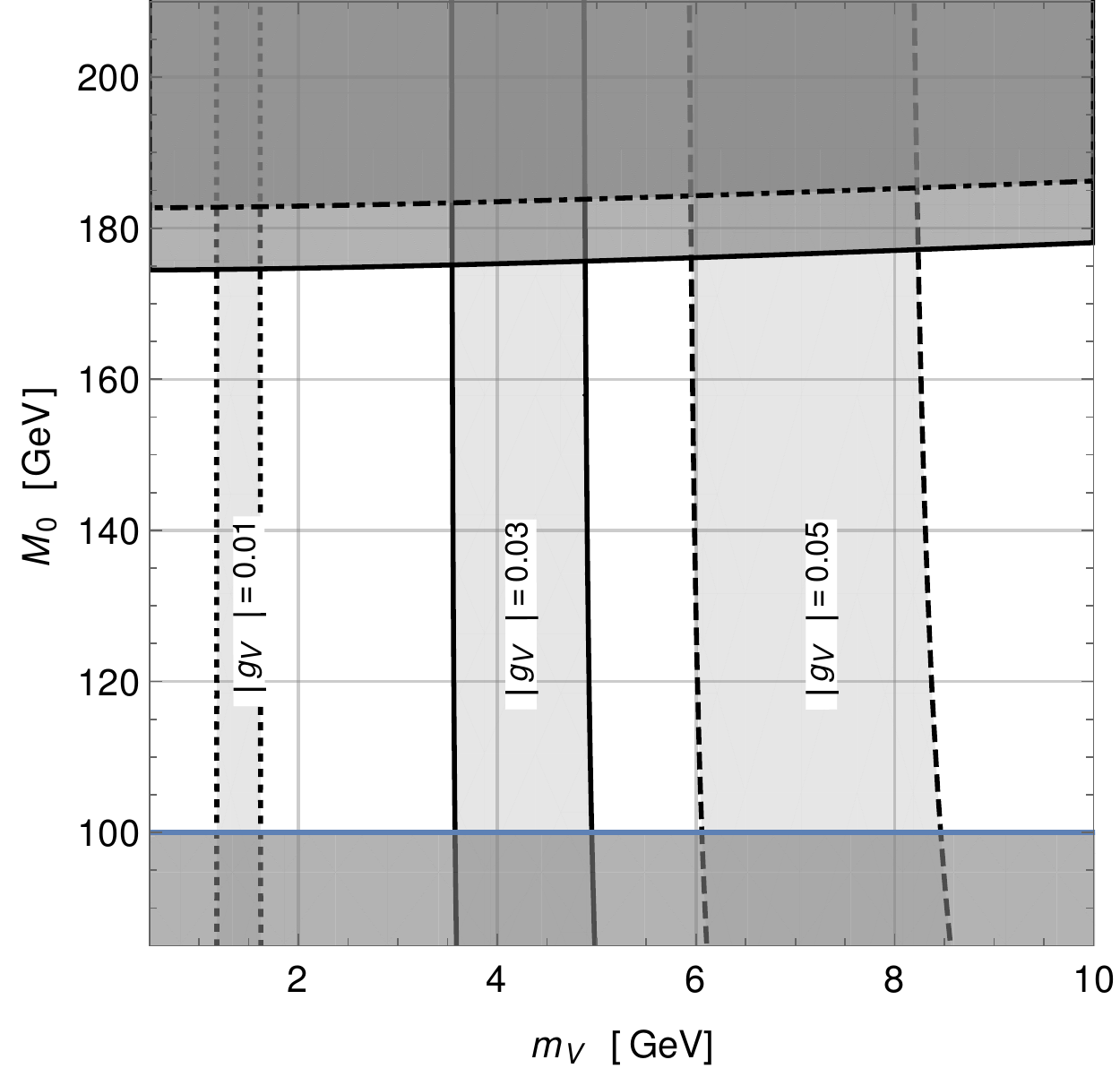} \quad
 	\includegraphics[width=0.45\textwidth]{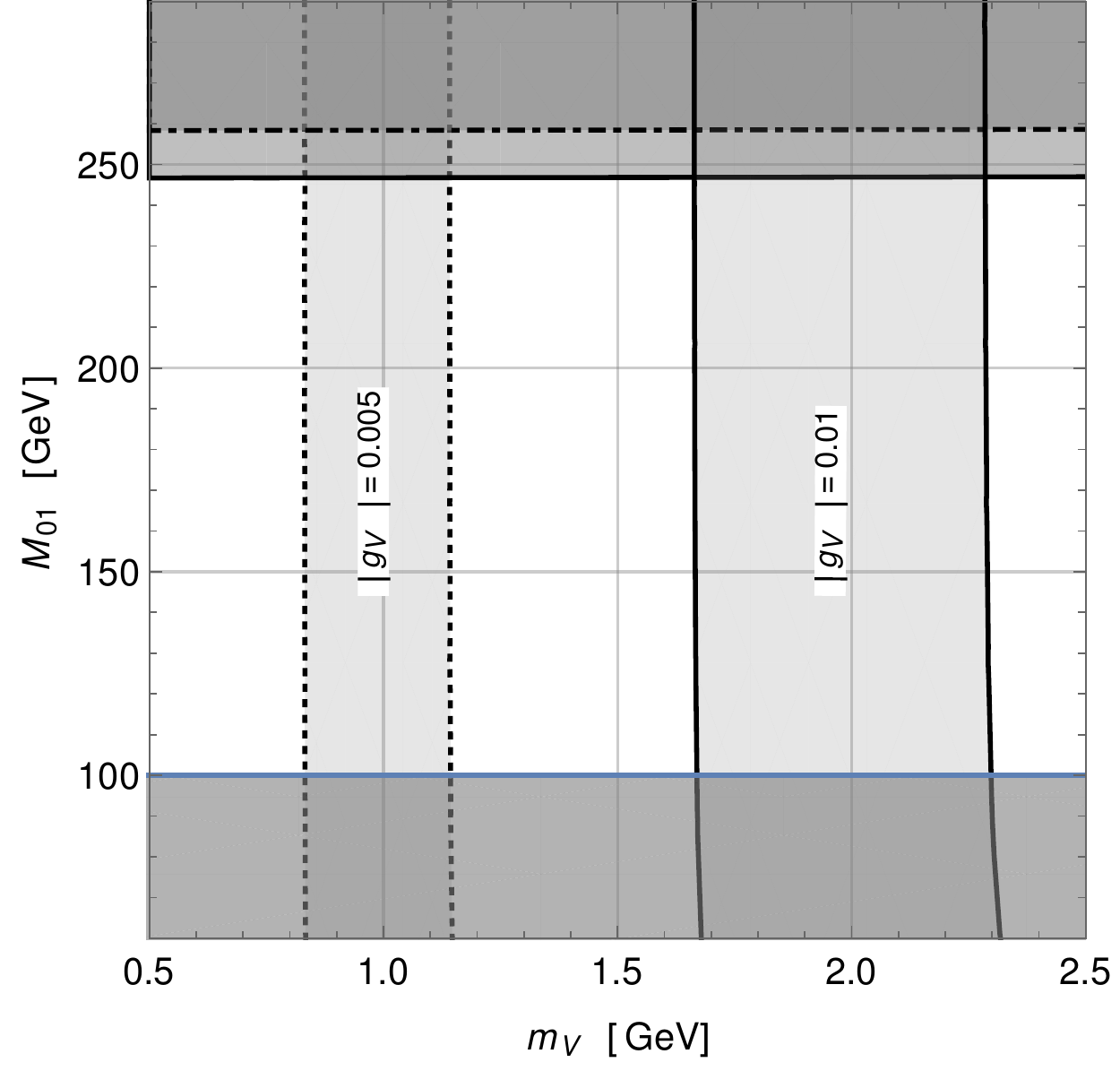}
	   \caption{\label{Fig:g_M_vec} Allowed regions in the vector meson-mumeson mass plane for 
	   the neutral mumeson(s) with flavor-universal LM interactions.  
	   The dark grey area in the bottom is covered by the LEP searches for new particles. 
	  The top dark grey areas with solid and dot-dashed boundaries are disfavored by 
	  LEP constraints on $ee\ell\ell$ contact interactions to explain $a_\mu$ (within 
	  the 1\,$\sigma$ range) for the left-chiral and right-chiral case, respectively. 
	   %
	 {\it Left}:  Case of a single mumeson $\mu_V^0$ with the coupling $g_V$, where the
	  light grey regions with dotted, solid, and dashed boundaries are allowed by the current 
	  $a_\mu$ data (within 1\,$\sigma$) for $|g_V|=0.01$, 0.03 and 0.05, respectively. 
	  %
	 {\it Right}:  Case of two mumeson-meson-lepton couplings $g_{V1}=\pm g_{V2}$, where the 
	  light grey regions with dotted and solid boundaries are allowed by  
	  $a_\mu$ data (within 1\,$\sigma$) and correspond to $g_V\equiv g_{V1}=\pm 5\times10^{-3}$ and $\pm0.01$, respectively. 
	 }
   \label{Figdistr}
 \end{figure}
\end{center}
\subsection{Vector leptomeson contribution}

For vector mumesons $\mu_V$ the value of $a_\mu$ is sensitive to the meson masses and almost 
insensitive to the mumeson masses. 
For one neutral mumeson interaction the allowed range of the mumeson mass values 
is ${100~\text{GeV}<M_0\lesssim180}$~GeV, 
which corresponds to the allowed area between the lower and upper dark grey bands 
in Fig.~\ref{Fig:g_M_vec} (left). The three light grey areas in this figure 
correspond to the ranges allowed by the current $a_\mu$ data (within $1\,\sigma$) 
for the chosen values of the coupling of $|g_V|=0.01$, 0.03, and 0.05, 
where $g_V\equiv g^0_{\alpha V}$ ($\alpha$ is either $L$ or $R$). 
Clearly only small values of $|g_V|=\mathcal{O}(10^{-2})$ are allowed.

The neutral leptomesons $\ell^0_V$ generate four-lepton contact interactions $ee\ell\ell$ 
through the box diagram shown in Fig.~\ref{Fig:box_V} (right). 
These interactions can be calculated and compared to the effective expression 
containing the contact interaction scale $\Lambda$ 
as~\cite{Freitas:2014pua,Kroha:1991mn}
\begin{eqnarray}\label{Eq:contact}
 \frac{(g_{V}^{e})^2(g_{V}^{\ell})^2}{64\pi^2m_{V^\pm}^2}F\left(\frac{M_0^2}{m_{V^\pm}^2}\right) 
 (\bar e_\alpha \gamma_\mu e_\alpha)(\bar\ell_\alpha \gamma^\mu \ell_\alpha) \leq \frac{4\pi}{(1+\delta)\Lambda^2} 
 (\bar e_\alpha \gamma_\mu e_\alpha)(\bar\ell_\alpha \gamma^\mu \ell_\alpha),
\end{eqnarray}
where we restored the flavor index of the couplings $g_V^\ell\equiv g_{\alpha V}^{0\ell}$ and assumed a common mass scale 
$M_0$ of the LMs $\ell_V^0$, 
the parameter $\delta=1\,(0)$ for $\ell=e$ ($\ell\neq e$), and the loop function can be written as
\begin{eqnarray}
 F(y) = \frac{1}{(y-1)^3}  [y^4 - 16y^3 + 19y^2 +2(3y^2+4y-4)y\,\text{ln}\,y-4]>0.
\end{eqnarray}
For the flavor-universal couplings $g_V\equiv g_V^\ell$ the limit in Eq.~\eqref{Eq:contact} can be rewritten 
using Eq.~\eqref{Eq:AB_V0_approx} as
\begin{eqnarray}\label{Eq:limit}
 \pi\, (\Delta a_\mu^\text{min})^2 \, \frac{m_{V^\pm}^2}{m_\mu^2} \, F\left(\frac{M_0^2}{m_{V^\pm}^2}\right) 
      f_\text{VVF}^{-2}\left( \frac{m_{V^\pm}^2}{M_0^2} \right)  \leq  \frac{m_\mu^2}{(1+\delta)\Lambda^2},
\end{eqnarray}
where $\Delta a_\mu^\text{min}$ is the minimal allowed value of $\Delta a_\mu$, e.g., within the $1\,\sigma$ range: 
$\Delta a_\mu^\text{min}=18.1\times10^{-10}$ from Eq.~\eqref{Eq:a_mu}. 
In the considered case of constructive interference between the SM process and the contact interactions, 
the constraints from the LEP measurements of $e^+e^-\to\ell^+\ell^-$ 
processes correspond to the lower limits of  
$\Lambda=12.7$~TeV for $\alpha=R$ and $\Lambda=13.3$~TeV for $\alpha=L$ (for $\ell=\mu,\tau$,
which gives stronger limits)~\cite{Alcaraz:2006mx}, 
and exclude parts of the parameter space, which are painted 
dark grey in the top of Fig.~\ref{Fig:g_M_vec} (left).

In the case of two neutral vector-mumeson contributions to $a_\mu$, which have the common LM and meson mass 
scales $M_0$ and $m_V$, the left-hand side of Eq.~\eqref{Eq:limit} gets an additional factor of
\begin{eqnarray}
  \frac{1+\rho^4}{(1+\rho^2)^2}  \geq  \frac{1}{2},
\end{eqnarray}
where $\rho$ is the ratio of the two couplings. The minimal value of 1/2 of this factor is achieved for the equal couplings 
(up to the sign) and corresponds to the weakest constraint from the contact interactions.
Figure~\ref{Fig:g_M_vec} (right) illustrates the case of either two vector mumesons with approximately 
equal masses $M_{01}\approx M_{02}$ interacting with 
the same charged meson or one mumeson with the mass $M_0\equiv M_{01}$, which significantly interacts 
with the two mesons with close masses $m_{V^\pm_1}\approx m_{V^\pm_2}$. 
In this figure we assumed equal meson-mumeson-lepton couplings $|g_{V1}|=|g_{V2}|$, 
where $g_{Vi}\equiv g^{0\ell}_{\alpha Vi}$ with fixed $\alpha$. 
The allowed range for the values of the mumeson mass  of ${100~\text{GeV}<M_{01}\lesssim250}$~GeV 
corresponds to the space between the lower and upper dark grey bands. 

\begin{center}
 \begin{figure}[tb]
 \centering
 	\includegraphics[width=0.45\textwidth]{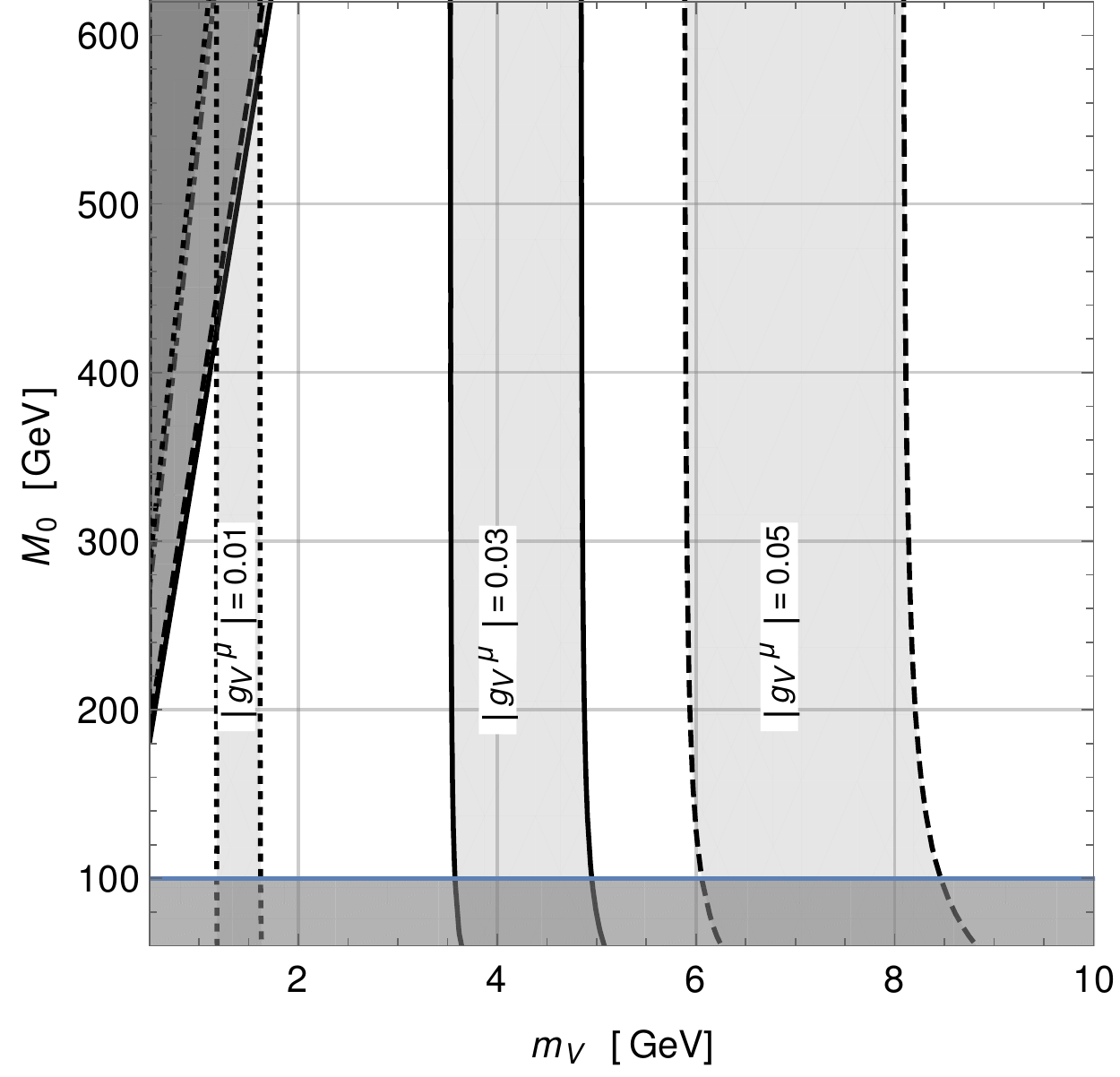} \quad
 	\includegraphics[width=0.45\textwidth]{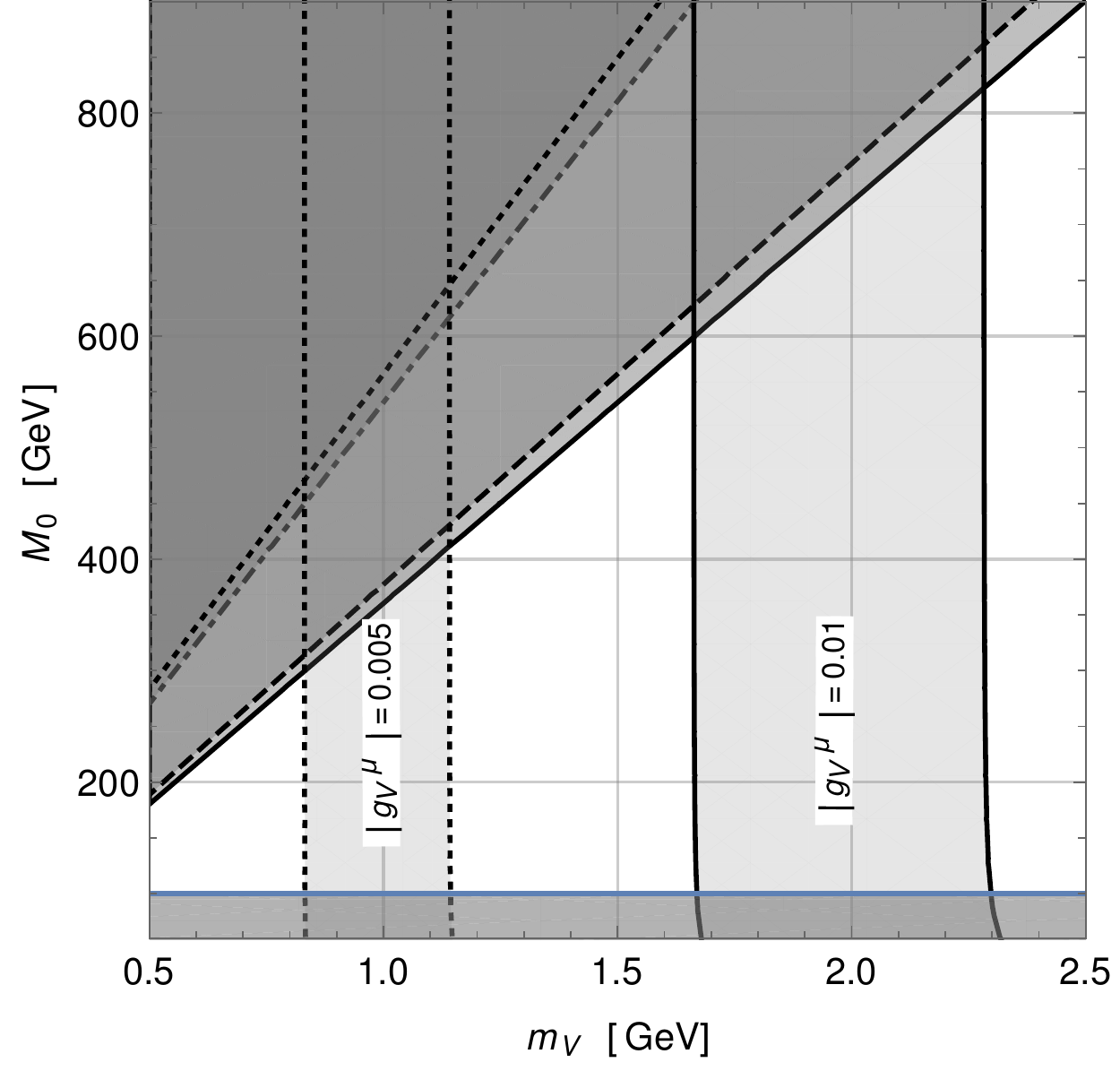}
	   \caption{\label{Fig:g_M_vec_nonFU} 
	   Same as Fig.~\ref{Fig:g_M_vec} but for the flavor-nonuniversal LM interactions. 
	   The dark grey regions with solid and dashed (dot-dashed and dotted) boundaries 
	   are disfavored by 
	  LEP constraints on the $ee\mu\mu$ contact interaction to explain $a_\mu$ within 
	  the 1\,$\sigma$ range for the left-chiral and right-chiral case, respectively, taking 
	  $|g_V^e|=3\times10^{-3}$ ($|g_V^e|=2\times10^{-3}$). 
	 }
   \label{Figdistr}
 \end{figure}
\end{center}

For flavor-nonuniversal couplings $g_V^e \neq g_V^\mu$ the limit in Eq.~\eqref{Eq:contact} can be rewritten as
\begin{eqnarray}
   \frac{(g_V^e)^2}{16\pi}  \, \Delta a_\mu^\text{min} \, F\left(\frac{M_0^2}{m_{V^\pm}^2}\right) 
      f_\text{VVF}^{-1}\left( \frac{m_{V^\pm}^2}{M_0^2} \right)  \leq  \frac{m_\mu^2}{\Lambda^2},
\end{eqnarray}
which is also valid for the two neutral vector-mumeson contributions to $a_\mu$ with 
the common LM and meson mass 
scales. Figure~\ref{Fig:g_M_vec_nonFU} shows that in this flavor-nonuniversal case the limits from 
the contact interaction $ee\mu\mu$ can be suppressed by a small value of the coupling $g_V^e$.

Concerning the LHC searches for $\mu_{V}^0$, in the case of a SM gauge singlet, Drell-Yan production of singlet pairs 
may be not possible. Then cascade decays from heavier charged particles can be considered~\cite{Freitas:2014pua}.

Notice that in the case of long-lived neutral LMs, which can escape the detector,  
the generic LEP lower bound for the new particle masses of 100~GeV may be weakened for their masses.\footnote{
Some questions of models with very light leptohadrons with masses of $\mathcal{O}(1)$~MeV 
(which we do not consider in this paper) were discussed in Refs.~\cite{Pitkanen:1990,Pitkanen:1992}.}

\section{Conclusion}\label{sec:conclusion}

We have found the regions for LM model parameters which are allowed by the muon $g-2$ data 
and the LEP data. 
We considered various minimal models 
and have shown that the scenarios with one (two) scalar meson-LM-lepton interactions 
limit the value of the lightest charged LM mass from above by 280 (400) GeV. 
However the case of vector meson-LM-lepton interactions is, in general, less predictive. 

This is very useful to investigate further collider restrictions on LM model parameters. 
In particular, the charged scalar LM interactions can potentially be either strongly bounded 
or even ruled out through the searches for their Drell-Yan production at the LHC. 
Future collider experiments such as the ILC and the FCC 
have great prospects to probe a significant part of the leptomeson (leptodiquark) model parameter space, 
which is still allowed.\footnote{In case of discovery of LMs the question of 
utility of lepton-meson and meson-meson colliders will be opened.}

\section*{Acknowledgements} 
The author thanks 
the listeners of the seminar of the Division of Field Theory 
and Elementary Particles of the University of Silesia for useful comments. 
This work was supported in part by the Polish National Science Centre, grant number DEC-2012/07/B/ST2/03867, and 
German Research Foundation DFG under Contract No. Collaborative Research Center CRC-1044. The author used 
JaxoDraw~\cite{Binosi:2003yf} to draw the Feynman diagrams.

\end{document}